\documentstyle[preprint,aps,prc]{revtex}
\begin{document}
\draft
\bibliographystyle{prsty}
\title{Splitting of the Dipole and Spin-Dipole Resonances}
\author{Sam M.~Austin$^1$\thanks{Electronic address:  austin@nscl.msu.edu},
E.~Adamides$^1$, A.~Galonsky$^1$, T.~Nees$^1$,
W.A.~Sterrenburg$^1$\thanks{Present address: Dept. of Physics,
Free University, 1007 MC Amsterdam, The Netherlands} ,
D.E.~Bainum$^2$\thanks{Present address: Washburn University,
Topeka KS 66621}, J.~Rapaport$^3$,
E.~Sugarbaker$^4$\thanks{Present address: Van de Graaff Lab., Ohio
State University, Columbus, OH 43212}, C.C.~Foster$^5$,
C.D.~Goodman$^5$, D.J.~Horen$^6$\thanks{Present address: 19
Colonial Ave., Framingham, MA 01701},
C.A.~Goulding$^7$\thanks{Present address: LANL, MS J562 NIS-6, Los
Alamos, New Mexico 87545}, and M.B.~Greenfield$^7$\thanks{Present
address:  Department of Physics, Int'l Christian University,
Mitaka, Tokyo 181 Japan}}

\address{$^1$ National Superconducting
Cyclotron Laboratory and Department of Physics and Astronomy \\
Michigan State University, East Lansing, MI~48824\\  $^2$
Department of Physics, Emporia State University, Emporia,
KS~66801\\ $^3$ Department of Physics, Ohio University, Athens,
OH~45701\\ $^4$ Department of Physics, University of Colorado,
Boulder, CO~80309\\  $^5$ Indiana University Cyclotron Facility,
Indiana University,  Bloomington, IN~47405\\ $^6$ Oak Ridge
National Laboratory, Oak Ridge, TN~37830\\$^7$Department of
Physics, Florida A\&M University, Tallahassee, FL32307}

\date{\today}
\maketitle
\begin{abstract}
Cross sections for
the $^{90,92,94}$Zr($p,n$) reactions were measured at energies of
79.2 and 119.4~MeV.  A phenomenological  model was developed to
describe the variation with bombarding  energy of the position of
the $L=1$ peak observed in these and other ($p,n$) reactions.  The
model yields the splitting between the giant
dipole and giant spin dipole resonances. Values of these
splittings are obtained for isotopes of Zr and Sn  and for
$^{208}$Pb.
\end{abstract}


\section{Introduction}
Inelastic scattering and charge exchange
reactions induced by intense neutrino fluxes during supernova
explosions play an important role in nucleosynthesis in
supernovae.  For example, they may produce rare light isotopes
such as $^7$Li, $^{11}$B and $^{19}$F\cite{Woosley90} that are not
sufficiently produced by the usual processes of nucleosynthesis
and may modify the distribution of elements formed during the
r-process\cite{Haxton97}. Observation of supernova neutrinos in
terrestrial detectors sensitive to neutrino flavor\cite{Fuller99}
may allow one to determine or limit the masses of the $\mu $ and
$\tau $ neutrinos and to determine whether oscillations occur.
From the point of view of nuclear physics, all these phenomena
share a common ingredient:  they depend on the $L=0$ allowed
(Gamow Teller) and $L=1$ forbidden strengths of transitions
induced by the various neutrino flavors.  It may be possible to
measure some of these strengths directly using neutrino beams, but
it will be necessary to infer most of them from the cross sections
for hadronic charge exchange and inelastic scattering reactions.

Unfortunately, information on the nature of the $L=1$ strength is
not yet sufficiently detailed for such applications. The spectrum
observed at low momentum transfer in a typical ($p,n$) reaction is
dominated by three isovector excitations:  the isobaric analog
state (IAS), the giant Gamow Teller Resonance (GTR), and the giant
dipole ($L=1$) resonance\cite{osterfeld92,Rapaport94,Alford98}. It is well
known that the IAS contains almost all the Fermi strength and that
the Gamow Teller peak contains about 60\% of the sum rule value.
Much less is known about the dipole peak and it exhibits a rather
unusual behavior; it moves to lower excitation energy as the
bombarding energy increases as shown in Fig. 1.

The qualitative reason for this
behavior seems clear; the $L=1$  peak contains two separate
structures: the giant dipole resonance (GDR) formed by a spin
transfer of $S=0$,  and the spin  dipole resonance (SDR) formed by
$S=1$. These resonances lie at different positions, but are
sufficiently wide that they overlap. Because their relative
excitation is expected to change in the energy range from 40 to
200 MeV, that of the SDR becoming relatively stronger at higher
energy\cite{Osterfeld81}, the $L=1$ peak will shift toward the SDR
as the energy increases. The amount of the shift depends on the
energy splitting of the GDR and SDR;  it should then be possible
to obtain an estimate of this splitting from an analysis of the
observed shifts.

The Zr isotopes are an excellent place to study the shift
phenomenon. There have been ($p,n$) studies on
$^{90}$Zr\cite{Nishihara85,Sterrenburg80,Bainum80,Horen81} at 41,
45, 120, and 200 MeV, and on  other Zr isotopes at 45
MeV\cite{Sterrenburg80} and 200 MeV\cite{Horen81}. This paper
describes measurements of the $^{90,92,94}$Zr($p,n$) reactions at
two intermediate energies, 79.2 and 119.4 MeV; a model that can
account for the energy variation observed for the most  complete
data set ($^{90}$Zr); and  the application of this model to
extract the splitting of the $S=0$ and $S=1$ components of the
$L=1$ resonance for several other nuclei.  The basic assumptions
of this model are similar to those employed by Taddeucci {\em et
al.}\cite{Taddeucci87} in the description of  $L=0$ charge
exchange reactions with $S=0$ and $S=1$.

The model also provides an improved estimate of the position of
the GDR obtained from ($p,n$) reaction studies.
In earlier work  it has simply been assumed that at bombarding
energies near 50 ~MeV  the observed excitation is dominated by the
GDR; we show that this  is not the case. We note that one cannot simply
obtain the position of the GDR in the charge exchange channel from
photonuclear studies on the target nucleus.  In nuclei with substantial
values of N-Z, the dominant excitation in ($p,n$) is the $T-1$ component of the
GDR where T is the isospin of the target nucleus.  The collective shift of
this resonance differs from those of the
$T$ and $T+1$ components seen in the target.  Phenomenologically, this
is accounted for by an isotensor component of the effective isospin
potential\cite{Sterrenburg80,leonardi72}, which is poorly known.  Consequently,
the location of the $T-1$ GDR must be taken as unknown.

\section{Experimental Procedures and Results}
Data were taken using the beam swinger time of flight facility at
the Indiana  University Cyclotron Facility\cite{Goodman79}, with
two 15~cm $\times$ 15~cm $\times$ 100~cm detectors, located 46.5~m
and 70.15~m from the target.  Bombarding energies were 79.2~MeV
and 119.4~MeV, nominally 80 and 120 MeV.  Targets were enriched
$^{90,92,94}$Zr metal with thicknesses of 77.5, 25.4, and 41.6
mg/cm$^2$, respectively. The overall resolution was between 415
and 540 keV (580 and 675 keV) at  79.2 MeV (119.4 MeV),
significantly smaller than the width of the $L=1$ peaks we are
concerned with here. Spectra were taken at about 0.1, 4, 8, 12, 16
and 19.5 deg(lab). Those for $^{90}$Zr at 4 deg are shown in Fig.
2; at this angle the IAS and the  GT resonances,  which peak at 0
deg, and the $L=1$ resonances, which peak at 7--8 deg  are
visible.

Spectra were fitted with sums of Gaussian peaks plus a quadratic
background, as shown in Fig. 2. The resulting
fits are satisfactory. For angles of 4 deg and greater a single
gaussian  provides a good description of the $L=1$ strength.  This
is somewhat surprising because the $S=0, S=1$ splitting is
expected to be in the range of 3.0 to 4.5 MeV\cite{Krmpotic83},
not so much smaller than the width of the GDR, and because the
relative contributions of the GDR and SDR are expected to change
from 1:1 to about 1:20 over the energy range studied.

A numerical experiment was performed to evaluate the uncertainties
that might arise from this procedure.  Gaussian peaks with widths
of 4.5 MeV (representing the GDR) and 8 MeV (representing the
SDR), and 3.9 MeV apart, were added and their sum was fitted by a
single Gaussian. This was done for peak ratios from 1:1 to 1:20.
In no case did the centroid of the fitted Gaussian differ from the
actual centroid by more that 0.15 MeV, and most differences were
much smaller. In another test, the centroid of the counts above
the fitted background was calculated and compared to the centroid
of the fitted Gaussian.  The difference was 0.17 MeV. This gives
one some confidence in the procedure followed at about the $\pm $
0.1 MeV level. The resulting energies for the dipole peak are
shown in Fig.~1; the $L=1$ peak shifts to lower energies by about
2 MeV as the bombarding energy changes from 40 to 200 MeV. The
peak positions are consistent with centroid energies obtained by others
at both lower and higher bombarding energies (see Fig. 4), lending some
confidence to the procedures used here.

\section{Model and Results}
We outline here a phenomenological model that describes the
observed shift with bombarding energy in terms of the positions of
the GDR and the SDR. The basic idea is as outlined in the
Introduction: that the energy of the $L=1$ peak moves toward the
energy of the SDR as the bombarding energy increases.  The
notation we use is shown in Fig. 3. We denote the energy of the
$S=0$ component, the $J^\pi$ = 1$^-$ GDR, by $E_0$. Similarly, the
energy of the $S=1$ component, the SDR consisting of 0$^-$, 1$^-$,
and 2$^-$ states has a combined centroid denoted by $E_1$. The
states with different $J$ will have different distributions of
strength with excitation energy.  We lack the information to treat
these differences in excitation in more detail, but believe that the
effects are probably small compared to other uncertainties in the model.

Because of the nature of the two-nucleon interaction mediating the
transitions, the strength of the $S=0$ excitation of the GDR is
comparable to that of the $S=1$ excitation of the SDR at 45 MeV,
but $S=1$ becomes much stronger than $S=0$ as the bombarding
energy increases toward 200 MeV. One then expects the observed
centroid of the $L=1$ strength to move toward  $E_1$ with
increasing bombarding energy. Rather general
arguments\cite{Krmpotic83} indicate that the $S=0$ GDR lies at
higher excitation energy than the centroid of the $S=1$ SDR so
this motion corresponds to a decreasing excitation energy.
Detailed theoretical calculations\cite{Osterfeld81} qualitatively
confirm this expectation.  See Refs.\cite{osterfeld92,Rapaport94,Alford98}
for general discussion and references.

The position, $C$, of the centroid of the $L=1$ excitations
(including both the GDR and SDR) at a bombarding energy $E_p$ is
given by
\begin{equation}
{C} = \frac{\sigma_0{
E}_0 + \sigma_1{E}_1}{\sigma_0 + \sigma_1} = { E}_0 -
\frac{\sigma_1/\sigma_0}{1+\sigma_1/\sigma_0}\,\, \Delta\,\, ,
\end{equation}
\noindent where $\Delta$=$E_0$ - $E_1$ and $\sigma_0(\sigma_1)$ is
the cross section for $S=0(S=1)$ transfer.  One can express $C$ in
terms of $E_0$ and $\Delta$ if one knows the ratio
$\sigma_1/\sigma_0$.  We assume here that for the $L=1$
excitations this ratio is directly proportional to the ratio of
the forbidden $S=1$ and $S=0$ beta decay strengths, as is observed
for the $L=0$ case\cite{Taddeucci87} where the analogous
excitations are the Gamow Teller and IAS resonances. This
assumption is confirmed for $L = 1$ by the DWIA calculations of
Gaarde {\em et al.}\cite{Gaarde81}. Then for small momentum
transfers one has~\cite{Taddeucci87}

\begin{equation}
\sigma_1/\sigma_0 \approx
\frac{{K}_{\sigma\tau}}{{K}_\tau} {N}_D^1 \left| \frac{{
J}_{\sigma\tau}}{{J}_\tau}\right|^2 \frac{B{\rm (SDR)}} {B\rm
(GDR)}\,\, ,
\end{equation}

\noindent where $J_{\sigma\tau}$ and $J_\tau$ are the volume
integrals of the parts of the effective interaction mediating
$S=1$ and $S=0$ transitions respectively, the $B$'s are the
corresponding transition strengths and the $K$'s are kinematic
factors whose ratio is very near 1.0 for our case. $N_D^1$ =
($N_{\sigma\tau}$/$N_\tau)_{{L}=1}$ where the $N$'s are distortion
factors which differ from each other by about 15\% at the first
peak in the $L=1$ angular distribution.  For $L=0$ transitions
${N}_D^0|{J}_{\sigma\tau}/ {J}_\tau|^2$ = $a^2E_p^2$ where $N_D^0$
= ($N_{\sigma\tau}$/$N_\tau)_{{L}=0}$,  $a$ = (55~MeV)$^{-1}$ and
$E_p$ is the bombarding energy\cite{Taddeucci87}. This
relationship is accurate over the energy range of 45 to 200~MeV
studied here. One then obtains

\begin{equation}
{C} = {E}_0 -
\frac{\alpha a^2 {E}_p^2}{1+\alpha a^2 {E}_p^2}\,\, \Delta \,\,.
\end{equation}

Here $\alpha$ is given by ($N_D^1$/$N_D^0$)[$B$(SDR)/$B$(GDR)]. In
principle, experimental results can be used to fix $\alpha$ if
data at three energies are available (there are three parameters
 $\Delta$, $E_0$ and $\alpha$). Fitting the
$^{90}$Zr results yields a value of $\alpha$ near 1, but with a
large uncertainty due to the limited accuracy of the centroid
data.  It is then desirable to obtain an estimate of $\alpha$ at
least partly from other considerations.

The ratio of the $B$'s depends on both the sum rule strengths for
the transitions and the fraction $f$ of those strengths (quenching
factor=$f$) found in the low excitation giant resonance peaks of
interest here.  Estimates with uncorrelated and with RPA ground
states\cite{Nakayamapc} give $B$(SDR)/$B$(GDR) =
3.25 and 3.2, respectively. Gaarde, {\em et al.}\cite{Gaarde81}
have also shown that the effect of RPA correlations is small. For
the $S=1$ quenching factor we take $f \approx$ 0.5, comparable to
that obtained experimentally for the $L=0$  GT
resonance\cite{osterfeld92}. A value of $f$ in the range of 0.3
to 0.5 has been observed by Gaarde {\em et al.}\cite{Gaarde81} for
masses of 40 to 208. Several authors, e.g. Dro\.{z}d\.{z} {\em et
al.}\cite{Drozdz87}, have shown that perhaps 30\% of the total
$L=1$ strength is moved to high energies. Strength at higher
excitation has also been seen in polarization transfer experiments and
in multipole decompositions of ($p,n$) and $(n,p$) data\cite{osterfeld92,Alford98}.
Our results then apply to the resonant part of the cross section,
not that part lying higher in the continuum. Implicitly the model assumes that
the largest concentration of strength, the peak in the excitation
energy distribution for $L=1$, gives a reasonable estimate of the
position of the collective strength for comparison with models. There
is some indication in Ref. \cite{Drozdz87} that the position of maximum strength
does not change greatly when continuum effects are considered.
For $S=0$ the quenching factor is presumed to be near 1.0, in analogy with the $L=0, S=0$
(Fermi transition) case.

Finally $N_D^1$/$N_D^0$ is obtained from
the DWIA calculations evaluated at the peaks of the relevant
angular distributions (e.g. near 4.5$^\circ$ for $L=1$ and at
0$^\circ$ for $L=0$ at 200~MeV on Pb) and typically has a value
near 1.0. Based on all these considerations we  take $\alpha$ =
1.5 $\pm$ 0.5 as covering the values of interest.

Fitting Eq.~3 with $\alpha$ = 1.5 $\pm$ 0.5 to the data for
$^{90}$Zr yields the results shown in Fig.~4. Similar procedures
were followed for other isotopes for which data are available at
widely separated bombarding energies. For $^{92,94}$Zr the
excitation energy data are shown in Fig.~1; for $^{112,116,124}$Sn
data are available at 45 and 200 MeV\cite{Sterrenburg80,Horen81};
and for $^{208}$Pb data are available at 41,45,120 and 200
MeV\cite{Nishihara85,Sterrenburg80,Horen81,Horen80}.  The results
are shown in Table~I. Also shown in Table~I is the position of the
GDR measured in pion induced charge exchange reactions. These data
are not in good agreement with the ($p,n$) results for $^{90}$Zr.
However, the difference between the results of single and double
charge exchange reactions for $^{208}$Pb indicates that there may
be systematic uncertainties not yet accounted for.

Krmpoti\'{c}, Nakayama and Pio~Gale\~{a}o\cite{Krmpotic83} and
Auerbach and Klein\cite{Auerbach83} have discussed the nature of
the giant first forbidden resonances and have given values for the
energies of the GDR and SDR:  their results are given in Table~I.
They are in qualitative agreement with the results of a detailed
calculation by Bertsch, Cha, and Toki\cite{Bertsch81} which places
the centroid of the SDR near 18~MeV, below the GDR.
Theory and experiment agree closely for $^{208}$Pb and
within the uncertainties for the Zr isotopes, although the theoretical
results are smaller.  The agreement is less good for the Sn isotopes,
perhaps because experimental data are available only at 45 and 200 MeV.
It is difficult to estimate the theoretical
uncertainties, but they could be a few MeV, mainly in the position of
the spin-dipole excitation\cite{Auerbach00}.

\section{Discussion and Summary}
It is perhaps surprising that the the experimental and theoretical
values of $E_0$ and $\Delta $ shown in Table I are in reasonably good agreement, since
the experimental results do not measure the position of the high-lying continuum
strength and the theoretical estimates include all strength. We do not have
an explanation for this result, except possibly that the position of
maximum strength calculated in RPA does not change much when continuum effects are
included; see Ref. \cite{Drozdz87}. Calculations of the positions of the
resonance peaks would be more appropriately compared with the present
experiments.

Information similar  to that in Table I
can in principle be obtained from a comparison
of the positions of the SDR observed in high energy ($p,n$) reactions
and the GDR observed in pion charge exchange; however, the energy
shift approach outlined here at the least provides an independent
measure of the GDR-SDR splitting, and arguably is less subject to
systematic uncertainties arising from different experimental
approaches. The present procedure is based on simple assumptions
and simple Gaussian shapes for the dipole resonances. The
procedure could be greatly improved if data of sufficient
statistics and angular range were available at several energies
for each target so that subtraction or multipole fitting
procedures could be used to isolate $L=1$ strength.  With better
data, it would become worthwhile to model corrections owing to the
(relatively small) contributions of the higher isospin ($T$)
components of the excitation;  here we have implicitly neglected
all but the $T = T_0 -1$ components.

Even in its  present form the model provides a quantitative
explanation of the shift of the $L=1$ excitation with bombarding
energy, a measure of splitting of the spin-dipole and dipole
resonances, and an improved estimate of the position of the GDR
peak (see Table~I and Fig.~4).  The results appear in a general
way to support the
assumptions made by Fuller {\it et al.}\cite{Fuller99} in their
calculations of the the response of a flavor-dependent supernova
neutrino detector. They assume that the spin-dipole resonance
does not lie at higher excitation energy than the GDR, in agreement
with our finding. However, the present results only apply to the
part of the strength that is in the resonant peaks, while the response
of the detector will depend also on the part of the strength that
lies in the continuum at higher $E_x$. In addition, it will be
necessary to use a model to relate the observed splitting in
the charge exchange channel to that in $^{208}$Pb.

The results also indicate that
past estimates of tensor contributions to the isospin splitting of
the giant dipole resonance (e.g. Sterrenburg {\em et
al.}\cite{Sterrenburg80}) must be modified.  These estimates
assumed that  the GDR was at the
position of the $L=1$ peak observed for bombarding energies near 50 MeV.
Our results show that it lies at higher $E_x$, reducing the values
of $E_{-} = E_T-E_{T-1}$ obtained in that paper.

To summarize, cross sections for the $^{90,92,94}$Zr($p,n$)
reaction have been measured at energies of 79.2 and 119.4 MeV. The
positions of the observed $L=1$ excitation have been extracted and
compared with those at other energies;  the excitation strength is
observed to move toward lower excitation energy as the bombarding
energy increases. A simple phenomenological picture was developed
to describe the observed energy variation and was used to extract
the splitting of the giant dipole (GDR) and giant spin-dipole
(SDR) resonances. The results are in agreement with theoretical
models as to the scale of the splitting, but the value of
$\Delta $ we obtain is usually  somewhat larger than predicted.

%\acknowledgments
 We have had valuable conversations with
N.~Auerbach, G.~Bertsch, D.M.~Brink, K.~Nakayama and H.~Toki.
E.~Adamides acknowledges
the financial support of the Academy of Athens--Greece from the
V.~Notara~Bequest. This research was supported by the National
Science Foundation and the Department of Energy.

\begin{figure}
\caption{Energies of the $L=1$ resonance for the Zr isotopes
measured from the IAS; values are from the present work, and from
Refs.\protect \cite{Nishihara85,Sterrenburg80,Bainum80,Horen81}.
The uncertainties of the present data include an estimate of the
systematic errors; this also appears to be the case for the data
from the other references.} \label{fig2}
\end{figure}

\begin{figure}
\caption{Spectra from $^{90}$Zr($p,n$) at 79.2 and 119.4~MeV, with
background used in the analysis. The energy is measured from the
IAS.} \label{fig1}
\end{figure}

\begin{figure}
\caption{Diagram showing assumed model and definition of terms.
The zero of the energy scale is the energy of the IAS.}
\label{fig3}
\end{figure}

\begin{figure}
\caption{Comparison of the $L=1$ resonance positions for $^{90}$Zr
and Eq.~3. Energies are measured from the IAS. The abscissa is the
bombarding energy and the curves are non-linear least square fits
to the data.} \label{fig4}
\end{figure}

\begin{table}
\caption{Energy\protect \tablenote{Energies are with respect to
the isobaric analog state} of the Dipole ($E_0$) and of the
Dipole--Spin-Dipole Splitting ($\Delta$)}
\begin{tabular}{cccccc}
&\multicolumn{2}{c}{Present\tablenote{Calculated for $\alpha$ =
1.5.  The uncertainties include an allowance for the uncertainty
of $\pm$ 0.5 in $\alpha$: one half the difference in $E_0$ and
$\Delta$ for fits with $\alpha$ = 1.0 and 2.0 is added linearly to
the fit error.  The first number following the $\pm$ sign is the
total error; the fit error is given in parentheses. }} &
\multicolumn{2}{c}{Theory\tablenote{From Ref. \cite{Krmpotic83},
except the second values of $E_0$ for $^{90}$Zr and $^{208}$Pb are
from Ref. \cite{Auerbach83}}}
 & $(\pi^+,\pi^0), (\pi^+, \pi^-)$\\ \tableline Target       &
 $E_0$        &  $\Delta$            & $E_0$ & $\Delta$   & $E_0$
 \\
               & MeV          & MeV                  &
 MeV           & MeV        & MeV \\ \hline $^{90}$Zr     &
 16.7$\pm$1.2(0.6) & 4.0$\pm$1.3(0.9)          & 15.3, 14.6          & 3.0 &
 14.2$\pm$0.5\tablenote{Mean of results in Refs. \cite{Erell86,Loveman89}} \\
 $^{92}$Zr     & 17.1$\pm$1.6(0.8) & 5.1$\pm$1.7(1.1)          & 15.0
 & 3.4  &
 \\ $^{94}$Zr     & 16.9$\pm$1.9(1.0) & 5.4$\pm$2.1(1.4) & 14.7          &
3.8         & \\ $^{112}$Sn    & 16.6$\pm$2.1(1.0) &
6.8$\pm$2.7(1.7) &
 13.8          & 2.8         & \\ $^{116}$Sn    & 17.5$\pm$2.5(1.2) &
 8.6$\pm$3.1(1.8)          & 13.1 & 3.3         & \\ $^{124}$Sn    &
 14.4$\pm$2.2(1.0) & 7.5$\pm$2.8(1.7) & 12.3          & 4.3         & \\
 $^{208}$Pb    & 11.0$\pm$1.5(0.8) & 4.7$\pm$2.0(1.4)          & 10.3, 9.9 & 4.7
 & 10.2$\pm$0.9\tablenote{Ref. \cite{Mordechai89}
  ($\pi^+, \pi^-)$}, 8.6$\pm$0.5\tablenote{Ref. \cite{Loveman89},
 ($\pi^+,\pi^0$)}\\
 \end{tabular}
 \end{table}

 
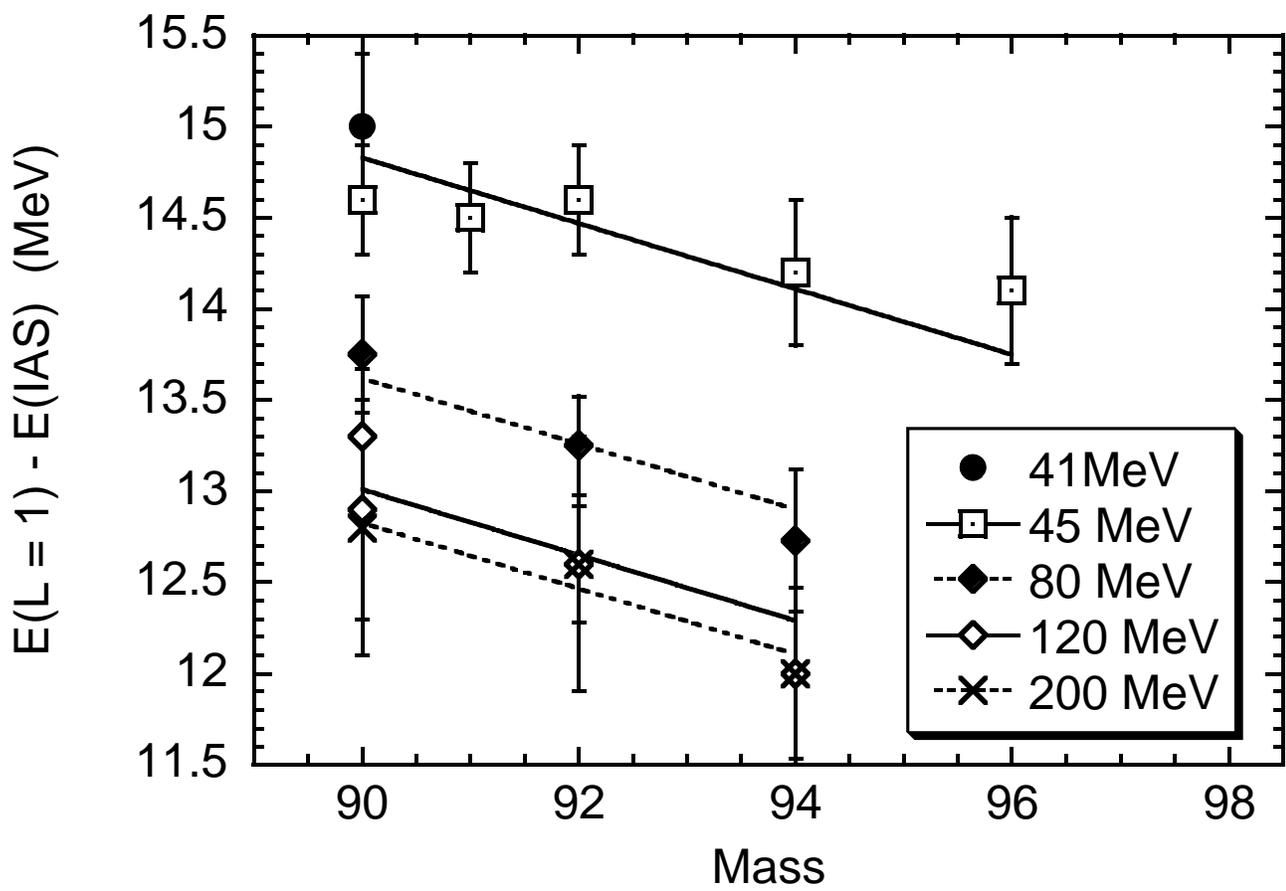
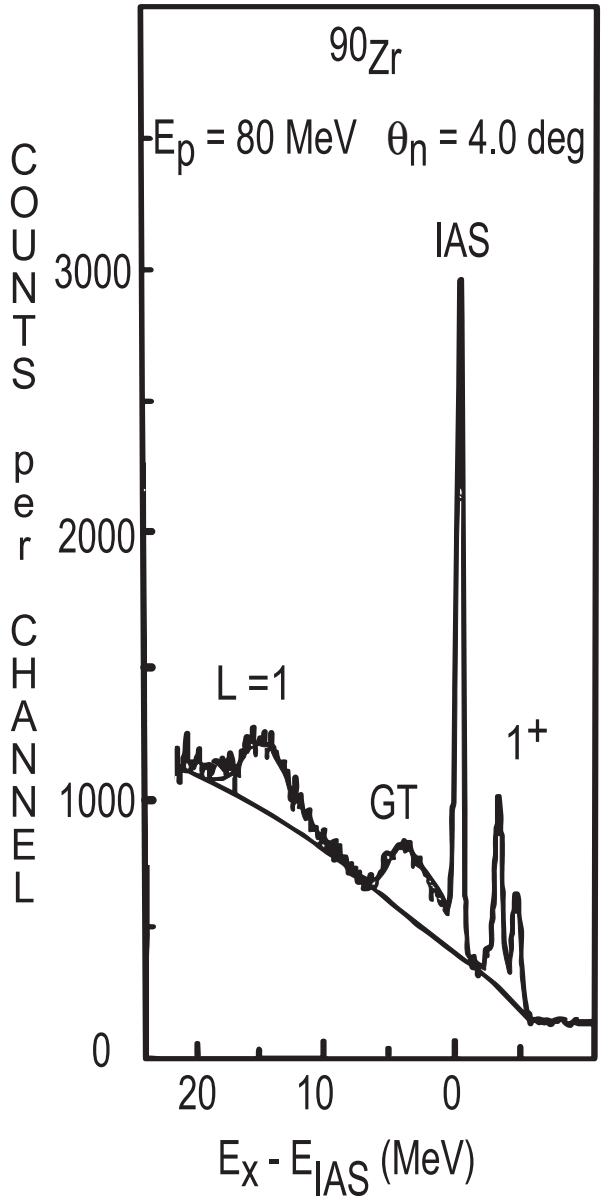
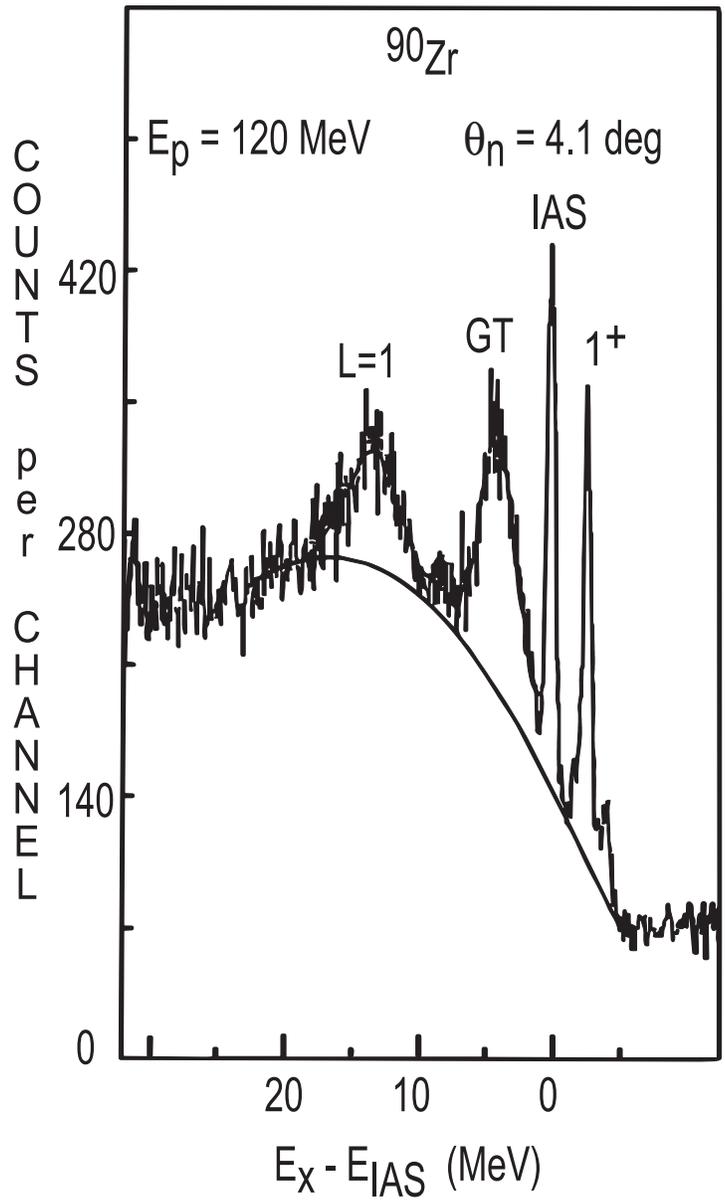
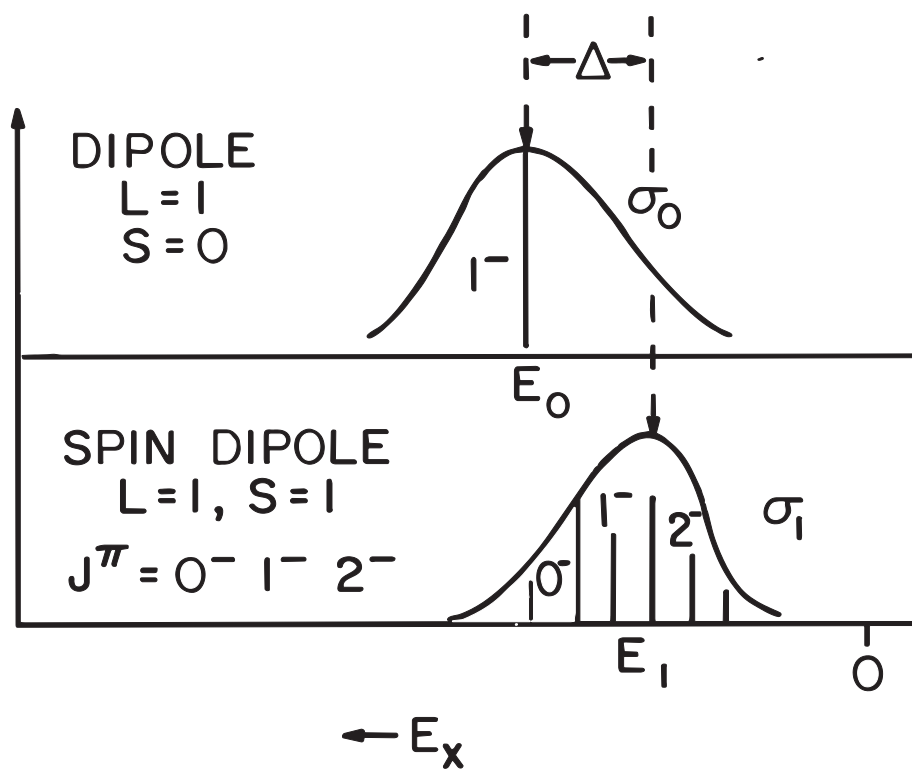
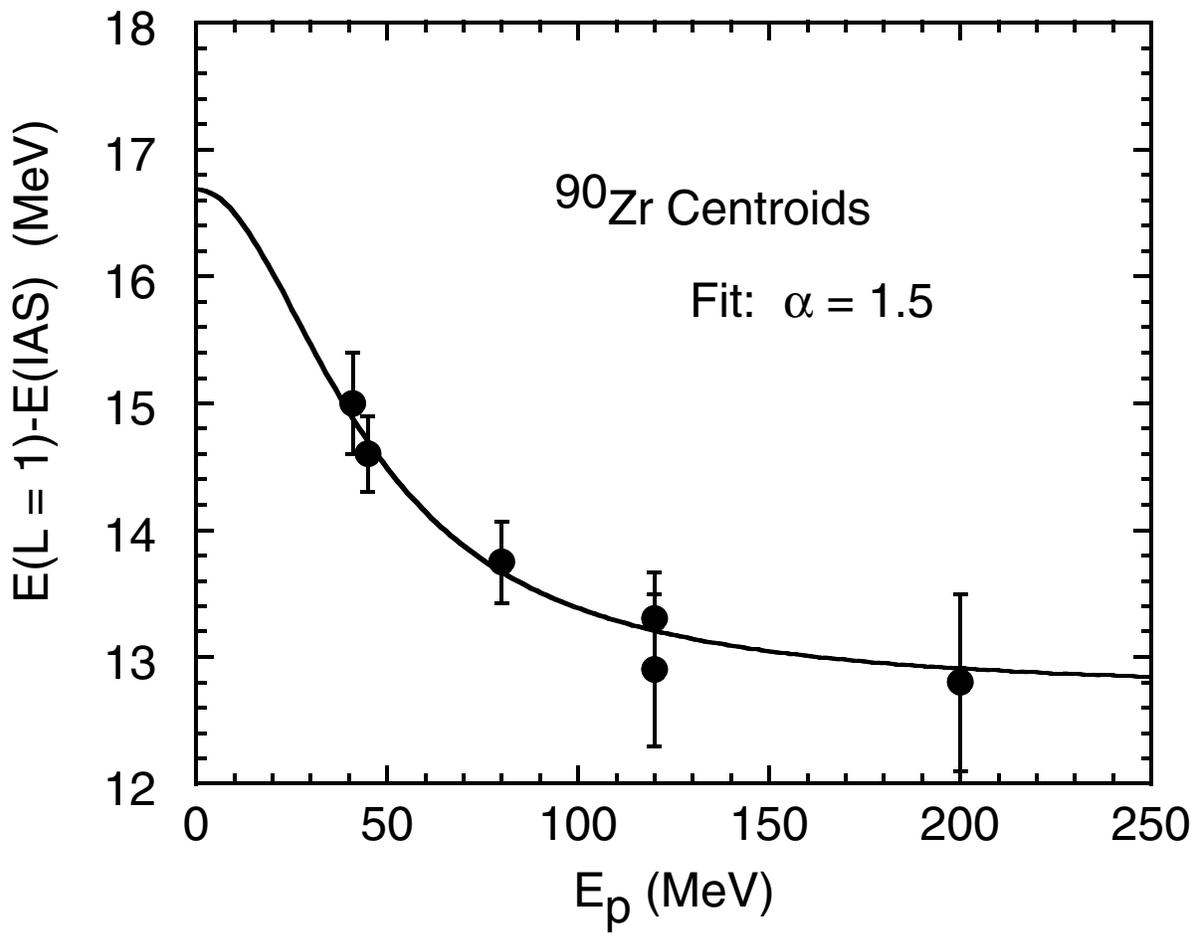
\end{document}